# Revisiting the City Tower Project: Geometric Principles and Structural Morphology in the Works of Louis I. Kahn and Anne Tyng

Aysan Mokhtarimousavi, Michael Kleiss, Mostafa Alani, Sida Dai

Clemson University, Clemson, SC
University of Maryland, College Park, MD
Tuskegee University, Tuskegee, AL
Virginia Tech, Blacksburg, VA

ABSTRACT: This paper presents a study of computation and morphology of Louis Kahn City Tower project. The City Tower is an unbuilt design by Louis I. Kahn and Anne Tyng that integrates form and structure using 3D space triangular geometries. Although never built, the City Tower geometrical framework anticipated later developments in design of space-frame structures. Initially envisioned in the 1950s the City Tower project is a skyscraper structure based on a tetrahedral and octahedral space frame called Octet-Truss. The aim of this study is to analyze the geometry of the City Tower structure and how it can be used to develop modular and adaptable architectural forms. The study is based on an analytical shape grammar that is used to recreate the original structure, and later to generate new structural configurations based on the City Tower's morphology. This study also investigates the potential applications of these findings in architecture and reveals the possibilities of using tetrahedrons and octahedrons as fundamental geometries for creating scalable and modular designs and presents initial findings.

**INTRODUCTION**
The collaboration between Louis Kahn and Anne Tyng is identifiable in several projects that the two architects completed together. The "genius for geometric form and topology" that Kahn has often claimed Tyng had been identified within the City Tower structure (Anderson, P. 2007). The City Tower project was designed for Philadelphia in the 1950s, a moment in architectural history when the integration between geometrical and structural innovation occurred. The City Tower project challenged conventional ideas about skyscrapers by using a space-frame system based on tetrahedral and octahedral geometries, emphasizing how form and structure become deeply integrated into each other (Juarez, A. unpublished). The tower's design reflects Tyng's research on geometric configurations and modular growth that sought to apply topological and organic principles to high-rise structures (Tyng, Anne 1975). Kahn and Tyng's work on this project illustrates the shift from traditional load-bearing systems to an interconnected triangulated space frame, allowing for adaptable spatial configurations (Juarez, 2022; Goldhagen, 2001). The City Tower is a fine example of geometrical and structural integration where architects sought to transcend conventional building forms, one that still has relevance more than 70 years after it was conceived. The City Tower first appeared in architectural discussions exploring how structure serves as the design's distinctive feature. Kahn and Tyng show how structure may go beyond practical requirements and be fundamental to create architectural identity and meaning. In this way, structural logic follows the evolution of tetrahedral-octahedral space-frame applications, similar to those explored by Buckminster Fuller in lightweight geodesic structures (Fuller, 1961), reinforcing its role as a seminal project in architectural geometry and structural innovation (Frampton, 1995)." The paper uses Shape Grammars to analyze the geometric principles of the City Tower structure, and shows how it be used can create modular, scalable frameworks.

**1.0 RESEARCH CONTEXT**
The postwar period was a transformative architectural era with the rise of new construction techniques and the search for a deeper architectural meaning beyond mere functionalism. One of the most influential figures in this period was Louis I. Kahn, whose work redefined the relationship between form, structure, and monumentality in architecture. Kahn's projects, particularly the Philadelphia City Tower and the Yale University Art Gallery showcased his belief that architecture should not merely serve as an enclosure but rather express its structural essence, integrating geometry, materiality and space into a unified whole (Hartoonian, Gevork. 2008). During this period, Kahn challenged the prevailing modernist dogma that emphasized lightweight, transparent structures, advocating instead for a return to monumentality and timelessness in architecture. His Philadelphia City Tower project (1952-1957) embodied this vision, as it was conceived as a triangulated space-frame structure—an innovative approach that emphasized the potential for self-supporting, modular growth in high-rise buildings (Doe, John 2010). Kahn's approach can be compared and related to two primary references. The first is a scientific study of topology through the influence of the French engineer Robert Le Ricolais. It has been suggested that "Kahn's engagement with the notion of 'growth' may have been influenced by D'Arcy W. Thompson's on Growth and Form, possibly introduced through his collaboration with Anne Tyng and her interest in geometric principles of form" (Juarez, A. unpublished). Anne Tyng was among the first women to receive an architectural degree (M. Arch.) and later worked for several years with Louis Kahn. She was strongly influenced by the work of Buckminster Fuller, which refers to applying geometric principles and techniques to the design and construction of buildings and structures (Tyng, Anne 1953). Bucky Fuller called "her discovery of Golden Mean relationships" between the whole family of Platonic Solids primary truths of ways to frame space (Anderson, P. 2007).

Topology, in both mathematics and architecture, focuses on spatial relationships and continuity rather than rigid geometric forms. This was central to Robert Le Ricolais' work. As a 'mathematician of structures, Le Ricolais was the first to introduce Topology as a tool for mathematical structural analysis (Sabini, M, 1985). In architecture, topology provides a flexible relational framework for understanding how spaces connect and interact, rather than just their fixed geometries and specific dimensions. Two figures are topologically equivalent when one can be obtained from the other by stretching or curving without cutting it. Topology is also the science of connectivity in the realm of form (Burry, J. and Burry M., 2010). This concept was fundamental to the City Tower where Kahn and Tyng explored the potential of tetrahedral and octahedral geometries to create a self-supporting adaptable spaceframe system. Their approach went beyond conventional structural logic to redefine the relationship between form, structure, and spatial hierarchy. The design integrated continuous spatial flow and hierarchical connectivity, reflecting topology's emphasis on spatial relationships over rigid predefined shapes (Juarez, A. unpublished).

**1.1 Octet truss structure**
The Octet Truss is a space frame system for lightweight rigid structures. Anne Tyng was familiar with the octet truss both as a product of Fuller's research and the work of Alexander Graham Bell Fig 1, whose octahedron-tetrahedron space structures and tetrahedral kites were often referenced in Tyng's publications (Tyng, Anne 1991).

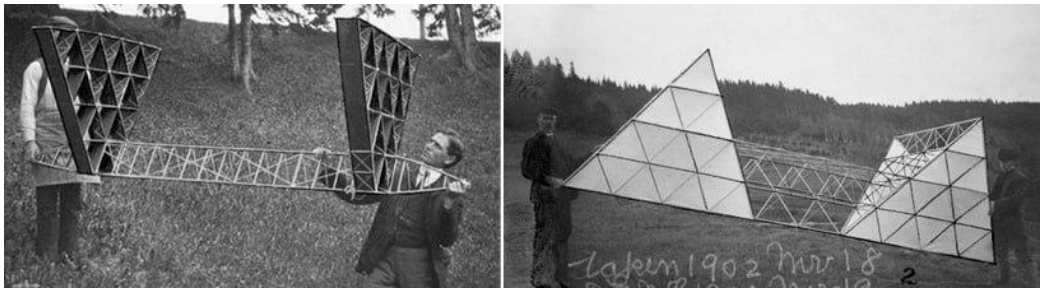

**Figure1:** Alexander Graham Bell Kite design. Source: www.messynessychic.com

Buckminster Fuller patented the octet truss in the 1960s as a system for 'Synergetic Building Construction.' The system is arranged by repeating a single spatial relationship between the octahedron and tetrahedron (Fig. 2). To achieve this relation; the two shapes maintain triangular faces of the same proportion so that they can align face-to-face as illustrated in Fig. 2(a) Further, the strength of the structural framework goes beyond that calculable by the known values of material strength alone. More precisely, the octet truss is a structure whose stress behavior is unpredicted by its parts. (Fuller, R. Buckminster. 1961)

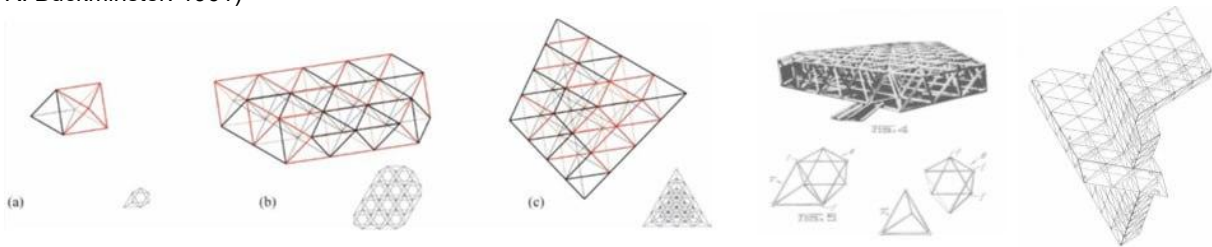

**Figure 2:** The Octet truss system, (a) typical octet truss parts and relations as defined by Buckminster Fuller; (b) sample octet truss assembly; (c) octet column (Ligler, H. 2023). (d), (e). Octet truss System depicted in 'Synergetic Building Construction' patent (Fuller, R. Buckminster. 1961).

The Octet-Truss geometry is one of the standard space-filling configurations (Fig 3) which features combinations of the simpler Platonic solids. The geometry is used as a repetitive module of octahedron-tetrahedron components in an arrangement that defines an octet truss.

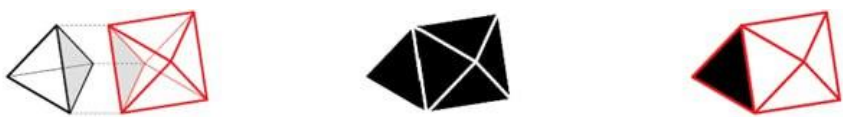

**Figure 3:** Octet configuration. **a**. wireframe view; **b**. hidden line view; **c**. color-coded hidden line view. Source: (Ligler, H. 2024)

EMERGING CHALLENGES
technological, environmental, social

## 2.0 THE STUDY

This study was divided into three phases. The first phase involves preliminary case studies where Kahn explored geometry (Elementary School and the Yale University art gallery). The second phase continues with an analysis of the City Tower geometry in the context of previous projects. The third phase concludes by developing the analytical Shape Grammar and exploring the implications of this computational tool for designing the city tower and other design configurations.

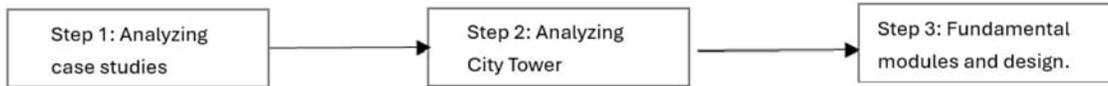

**Figure 4:** Research phases. Source (authors 2025)

### 2.1 Preliminary Case Studies: From Elementary School to City Tower

The proposal for an elementary school showcases a space-frame structural system beyond what Buckminster Fuller had already achieved. The design of a space-frame roof and its simultaneous use as a support element showcases a tectonic configuration dearting from the tectonic tradition of post and beam. This observation is present on Kahn's design for a tubular structure in Philadelphia City (Hartoonian, G. 2008) and was used later for designing City Tower. Anne Tyng embraced the ancient notion that the Platonic solids are the "primary truths of ways to frame space" (Tyng, Anne 1991). The geometry of the octahedron-tetrahedron system based on the octet truss later patented by R. Buckminster Fuller (1961) was a key element in the space-frame configuration. Tyng explored the octet truss as a continuous canopy and a columnar structure that emerges from the same geometry (Fig. 5). In the ceiling design, the truss is oriented so that triangles are organized in the horizontal plane. Concurrently with the development of this ceiling, Tyng focused on developing a transformation of the octet geometry that naturally supports the spaces with square rooms and a pitched roof (Burry, J. and Burry M., 2010)

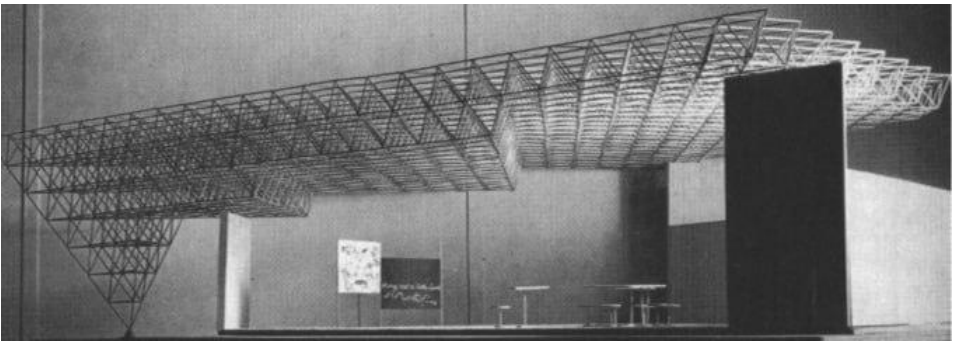

**Figure 5:** Ann Tyng, project for an elementary school, 1951. Source: (T. Leslie, Louis I. Kahn, 2005).

Tyng developed a layered triangular canopy composed of an octet framework supported by columns that continue the same geometry (Fig. 6). Tyng's octet-truss space structure functions in much the same way as Fuller's examples by defining a superstructure that contains the inhabitable volumes of the school beneath (Fuller, 1961). Tyng's design increases the layers of the tetrahedron/octahedron truss used by Fuller and Le Ricolais to 'grow' columns of the same geometry. The integral frame idea would be later developed in the City Tower project where the structure is understood as a continuous framework instead of as a cluster of separate elements. (Juarez, A. unpublished).

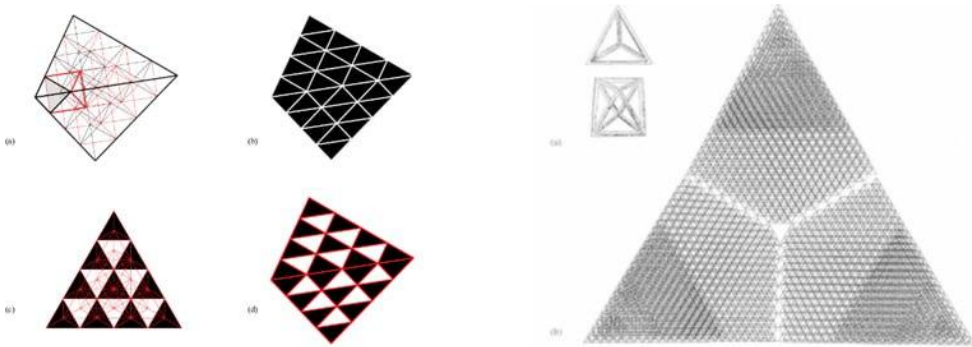

**Figure 6:** Elementary School (1949-51, Tyng). Octet column design from Proposed Elementary Schol. **a**. wireframe view; **b**. hidden line view **c**. plan view; **d**. color-coded hidden line view, (a) tetrahedron and octahedron components of space structure; (b) roof plan of ninety-six-foot structure. Source: (Tyng, Anne 1949)

**2.2 Yale University Art Gallery**
Many of the ideas for the City Tower also originated in the Yale University Art Gallery (1951-53). The absolute continuity of the geometrical order without any breaks was central to Kahn's thought. He demanded absolute structure integrity and insisted that a building clearly show how it was made and serviced. Moreover, he noted when he made that sketch that the structural order implicit in the tetrahedral floor slab of the recently finished Art Gallery should also have been extended into the columns "A tetrahedral concrete floor asks for a column of the same structure" (Heinz, R. and Sharad, J. 1987). The space structure geometry can be seen in the ceiling of the Yale Art Gallery (1951–53) (Fig 7). Here, Tyng collaborated with Louis Kahn to design a configuration in cast concrete oriented so that the base triangles of the tetrahedrons are open planes realized at the horizontal datum of the ceiling. In this orientation, the remaining three sides of each exposed tetrahedron are closed planes of concrete that conceal channels for ductwork and conduit within the structure (Fig. 7). (Ligler, H. 2024).

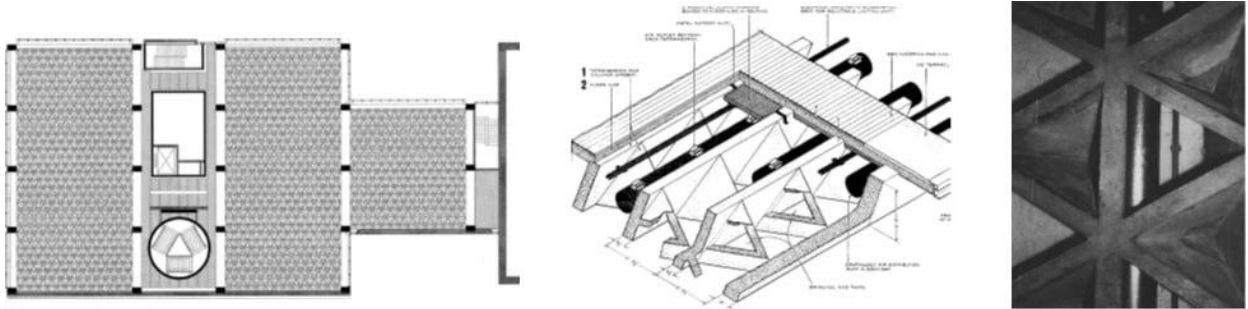

**Figure 7:** The Yale Gallery, A) Reflecting ceiling plan, B) section details, C) Detail of the ceiling. Source: (T. Leslie, I. Kahn, 2005)

In both precedents, the simple Platonic solids of the tetrahedron and octahedron are interrelated to show the architectural potential of these forms as spaceframes and void-defining structures. The integration of shape, form, geometry, and structure were explored. Tyng's mastery of the octet truss as a parametric and reconfigurable meta morphology provides a foundation on which future studies can build to show how these geometries reappear in other works produced along with Kahn. Most notably in the City Tower project where Kahn and Tyng derived forms by extending the triangular spaceframe vertically. (Ligler, H. 2023).

**3.1 ANALYZING THE CITY TOWER STRUCTURAL DESIGN**

The Concepts of natural growth and continuous structural unity are central to the City Tower. (Kahn, L. I. & Tyng, Anne. 1957). Kahn's interest in form and structure is magnified in this project which allowed him to reach beyond accepted dispositions for columns, beams, and walls (The Yale Architectural Journal, 1953). Kahn found that a geometrically ordered structure is the only way to integrate Form and Structure (Komendant, A 1975). The City Tower seems to reflect both: on one hand the octet-truss creates a tree-like structure as a whole; on the other, the connectivity between the geometric cells is esential to the spatial conception of the tower (Fig 8). The tower is an exercise in triangulation of structural members rising upward to form themselves into a vertical truss against wind forces.

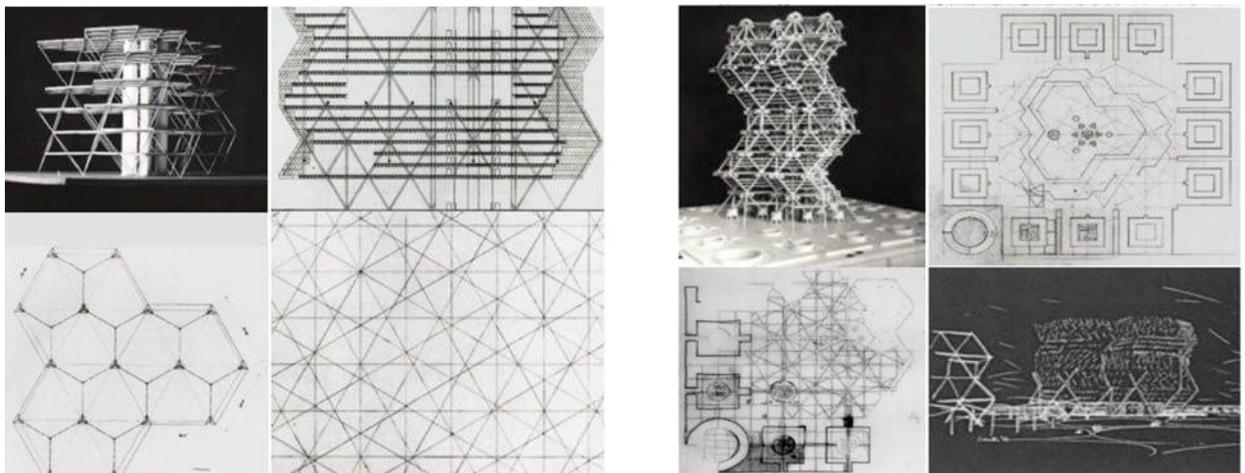

**Figure.8:** City Tower, Philadelphia, 1952-57, Source: (K. Frampton 1995).

The structure forms a triangulated braced spaceframe that intersects the system every 66 feet. Each intersection is crowned by an 11 feet deep space that houses storage, toilets, and substations for mechanical services. Intermediate-level floors (up to 6 within the 66-foot vertical bays) can be moved up and down within the triangulated envelope to suit specific planning and sectional requirements. No two floors align in plan. (Juarez, A. unpublished). The hexagonal plan of the structure rotates in vertical increments every foot. These undulating 66 feet shifts result from the natural completion of the triangulated spaceframe in its upward movement. The hierarchical expression occurs in variations in floor level between the primary 66 foot structural levels, in the hollow triangulated 'capitals,' high enough for a person to stand in, and the three-foot-deep hollow ceilings of octahedron-tetrahedron geometry. A central core of vertical shafts, which house stairs, elevators, and air ducts, rises through the building without disturbing its structural continuity (Tyng, Anne. 1983). Tyng showcases in the Philadelphia City Tower a representation of Platonic solids in a triangulated structure (Càndito, C., & Meloni, A. 2022). In the City Tower Tyng described the concept as one that 'extends the tetrahedron/octahedron building blocks vertically choosing directions of extension offered by the geometry' so that the standard octet truss (Fig. 9) is essentially reworked as a typical tower floor plate as illustrated in Fig. 9c. (Ligler, H. 2024).

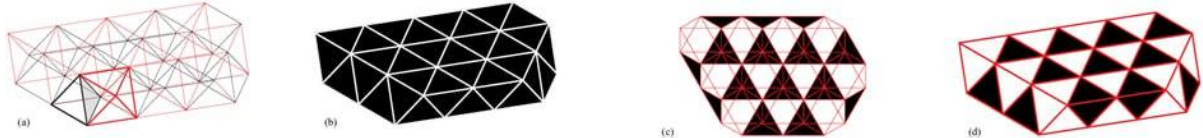

**Figure 9:** Octet truss system. **a**. wireframe view; **b**. hidden line view; **c**. plan view; **d**. color-coded hidden line View; Source: (Ligler, H. 2024)

**4.1 THE STUDY**
Based on analyzed case studies, and the Fuller octet truss configuration, the tetrahedron and the octahedron are interrelated and can be used to explore the potential for creating space-frame structural configurations. These two Platonic solids serve as foundational shapes for generating spatial configurations and designing a shape grammar for surface filling. Figure 10 illustrates A) the shape grammar, featuring two identical shapes (two tetrahedra and two octahedra), as well as one configuration with a tetrahedron and an octahedron combined, and B) examples of generated designs. (Kleiss et al. 2024)

**4.2 Shape Grammars**
A Shape Grammar is a production system that consists of four elements: 1) Shapes; 2) Spatial Relations; 3) Rules; and 4) Derivation (designs). The Shape Grammar determines the order in which these geometric operations occur because the grammar is not random. The output of the production system is a design or derivation that conforms to the rules and behaviors of the more extensive system or language from which it was derived. The theory of Shape Grammar has been applied in design, architectural, and computational fields of study for two purposes: 1) analyzing the design language behind a design creation and synthesizing the design language based on analyzing a known design language; and 2) creating new designs from the existing one by creating new rules or modifying the original rules. For Instance, in the study of Kleiss et al. (2024) two kinds of rules are considered: a one-shape/one-rule shape grammar for each of the two polyhedrons individually; and a two-shape/two-rule shape grammar using both polyhedrons, combining them into a more complex shape grammar and subsequent designs. The following figures show the spatial relations for each of the three shapes of the grammar and some of the generated designs.

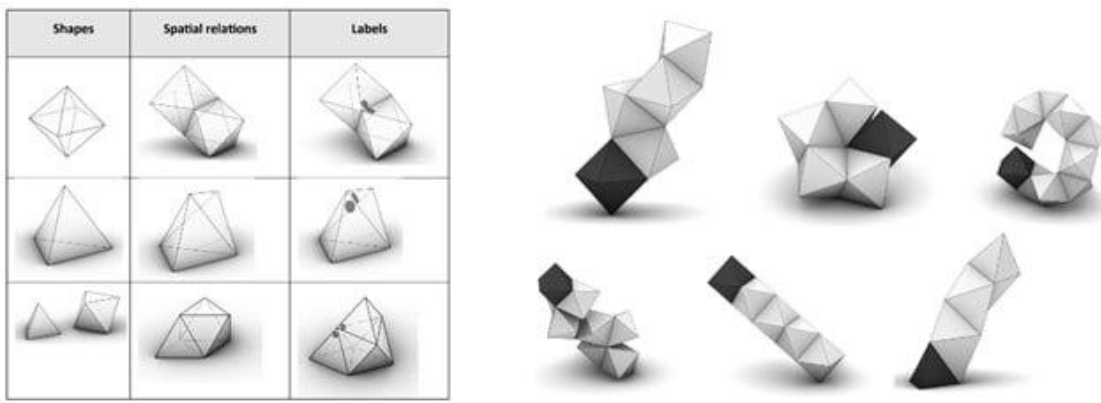

**Figure 10**: A) Spatial Relations Between Tetrahedrons and Octahedrons. B) Generated Designs from Shape Grammars. Source: (Kleiss et al. 2024)

This study creates hierarchical stages by dividing shapes into smaller geometries that make it possible to identify the starting point of the structure. Therefore, the starting point in Kahn's design process begins within its structural geometry which has been prepared in interdependent stages. The first stage considered two platonic shapes Tetrahedron and Octahedron as initial shapes. To achieve spatial relation between them, the two shapes maintain triangular faces of the same proportion to align face-to-face, edge-to-edge and vertex-to-vertex as spatial relations (Table 1). Further, two-shape/two-rule grammar using both polyhedrons are combined to generate fundamental modules of the City Tower project. It is important to note that due to the high symmetry of the two polyhedrons, many of the generated designs will not only be mirror designs but also identical with a different orientation in space. As a result, the combination of two initial shapes, Tetrahedron and Octahedron, is limited to three unique designs. These three configurations are considered foundation modules that will be examined as fundamental units for the city tower.

The study identifies which generated designs (illustrated in Table 1) can serve as the initial unit for the development of hexagonal plans to be vertically stacked to form a cohesive structural system for the entire building. To achieve this, the three possible units will be systematically analyzed to determine their suitability as an initial structural unit. The research methodology is structured into two key stages, focusing on the hierarchical organization and application of shape grammar: 1-Lower-Level Grammar: The lower-level grammar consists of rules for generating a "fundamental unit" design. The initial unit is analyzed as a lower-level grammar component for creating hexagonal shapes in this context. 2-Higher-Level Grammar: Higher-level grammar involves rules that generate the overall design through the recursive application of Euclidean transformations to the fundamental unit. The focus is on demonstrating the process of developing complex designs by applying simple, systematic rules at this higher level. The research investigates the grammar rules used to develop hexagonal plans.

**Table 1:** Tables should be aligned left on the page. Source: (Authors 2025)

| Design Numbers | Initial Shapes | Special Relation | Initial units (Lower-Level Grammar) | Half units | Hexagonal modules (Higher-Level Grammar) |
|---|---|---|---|---|---|
| 1 | 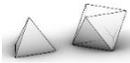 | 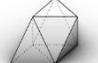 | 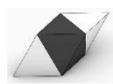 | 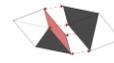 | 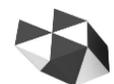 |
| 2 | 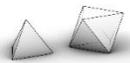 | 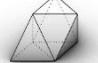 | 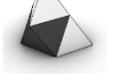 | 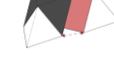 | 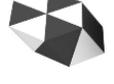 |
| 3 | 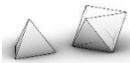 | 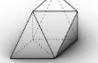 | 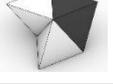 | 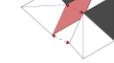 | 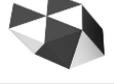 |

The design process for filling a 2D surface involves combining two tetrahedrons and one octahedron. These specific shapes are chosen due to their inherent geometrical compatibility, which results in a gapless arrangement when expanded along the X and Y axes. This compatibility provides a solid foundation for structured and modular design based on predefined geometric steps. Initially, the tetrahedrons contact the faces of the octahedron, forming a building block referred to as "Lower-Level Grammar." This lower-level grammar module establishes the base for creating large, cohesive structures. The configuration can then be conceptually halved; when the half-octahedron is combined with a tetrahedron, it results in smaller units called "half-modules." These half-modules are crucial in maintaining continuity and completeness in filling gaps within the overall structure. Geometric transformations are then applied to these modules to achieve hexagonal formation. The first module remains unchanged, while the second is rotated counterclockwise by 180 degrees and reflected. The half-modules are connected to form an infinite network of identical modules, creating a perfectly symmetrical and completely gapless pattern of hexagons. This configuration represents "Higher-Level Grammar," where recursive geometric rules allow the design to be extended and refined. Ultimately, the hexagonal module produced through this process becomes a highly effective building block. It can also be scaled to create more substantial architectural systems. Combining three of these modules forms a complete floor plan, which can be vertically expanded, showcasing both modularity and precision.

Table 2: Design process from conceptualizing initial units to developing the final design. Source: (Authors 2025)

| | Combination of Tetrahedrons to an Octahedron | Initial shapes generated by Lower-Level Grammar | Shape grammar Rules for connecting initial units | Generated designs | Connection of half-units to initial shapes | Generated design by Higher-Level Grammar | Connection Rules | Hexagonal Plan | City Tower final design |
|---|---|---|---|---|---|---|---|---|---|
| 1 | | | | | | | | | |
| 2 | | | | | | | | | |
| 3 | | | | | | | | | |

## 5.0 REFLEXION

The present study invokes the great potential of tetrahedrons and octahedrons as basic geometries for producing scalable and modulated architectures. Coherent patterns from the solids in question would quickly expand along the X and Y axes, using their direct geometrical complementarity and relatively high symmetry compared to the other Platonic solids. The research results in three alternate lower-level grammatical configurations upon applying a two-shape/two-rule grammar, which is fundamental to establishing higher-level grammatical configurations. This methodology on shapes allows for the generation of several designs by integrating these two shapes. These modules are scalable horizontally and vertically, forming the foundation for three-dimensional systems like the City Tower. The results highlight the potential of this approach to streamline fabrication and assembly, enhance structural performance, and promote environmentally responsible design and construction practices. (Mostafavi et al. 2024). The concepts explored in this study, particularly the use of tetrahedrons and octahedrons in Shape Grammar for modular and scalable design, provide a foundation for deployable structures that can be compactly folded and later expanded into their functional configurations, offering advantages in portability, adaptability, and ease of assembly. Given the geometric richness between tetrahedral and octahedral shapes, modules could be further extended across various scales, from small structural elements to large-scale space frames.

## 6.0 Acknowledgement

I would like to express my sincere gratitude to Farzad Saeidi Samet for his invaluable support in editing and refining the writing of this paper. His keen attention to detail and thoughtful feedback greatly contributed to the clarity and coherence of the final manuscript.